\documentclass[aps,pre,reprint]{revtex4-1}
\usepackage{amsmath}
\usepackage{graphicx}
\usepackage{color}
\usepackage{amssymb}
\usepackage{wasysym}
\usepackage{soul}
\input{epsf}

\begin{document}

\bibliographystyle{prsty}

\title{Polydispersity stabilizes Biaxial Nematic Liquid Crystals}

\author{S. Belli$^1$, A. Patti$^2$, M. Dijkstra$^3$, and R. van Roij$^1$}
\affiliation{$^1$Institute for Theoretical Physics, Utrecht University, Leuvenlaan 4, 3584 CE Utrecht, The Netherlands \\
$^2$Institute of Advanced Chemistry of Catalonia, CSIC, C/ Jordi Girona 18-26, 08034 Barcelona, Spain \\
$^3$Soft Condensed Matter Group, Debye Institute for NanoMaterials Science, Utrecht University, Princetonplein 5, 3584 CC Utrecht, The Netherlands}

\begin{abstract}

Inspired by the observations of a remarkably stable biaxial nematic phase [E.v.d. Pol {\it et al.}, Phys. Rev. Lett. {\bf 103}, 258301 (2009)], we investigate the effect of size polydispersity on the phase behavior of a suspension of boardlike particles. By means of Onsager theory within the restricted orientation (Zwanzig) model we show that polydispersity induces a novel topology in the phase diagram, with {\em two} Landau tetracritical points in between which {\em oblate} uniaxial nematic order is favored over the expected {\em prolate} order. Additionally, this phenomenon causes the opening of a huge stable biaxiality regime in between uniaxial nematic and smectic states.

\medskip

\noindent PACS numbers: 82.70.Dd, 61.30.Cz, 61.30.St, 64.70.M-

\end{abstract}

\maketitle

Since its first prediction back in the early 1970s \cite{freiser, alben, straley}, the biaxial nematic ($N_B$) phase has strongly attracted the interest of the liquid crystal (LC) community \cite{tschierske}. In contrast to the more common uniaxial nematic ($N_U$) phase, where cylindrical symmetry with respect to the nematic director determines optical uniaxiality, the $N_B$ phase is characterized by an orientational order along three directors and consequently by the existence of two distinct optical axes. The prospect of inducing orientational ordering along three directions, while maintaining a nematic fluid-like mechanical behavior \cite{berardi}, renders biaxial nematics preeminent candidates for next generation LC-based displays \cite{luckhurst}. Although experimental evidences of stable $N_B$ phases were reported already 30 years ago in lyotropic LCs \cite{yu}, in thermotropics this result was achieved in systems of bent-core molecules only a few years ago \cite{madsen}. Actually, when trying to experimentally reproduce an $N_B$ phase, one often encounters practical problems related to its unambiguous identification \cite{tschierske} and to the presence of competing thermodynamic structures \cite{taylor, vanroij, vanderkooij}. Stabilizing $N_B$ states is therefore an open, challenging scientific problem with huge potential applications. Motivated by the exciting results of a recent experiment on a colloidal suspension \cite{vandenpol}, we use here a mean-field theory to investigate the role played by size polydispersity on the stability of biaxial nematics in systems of boardlike particles. We show that a difference in the particle volume of a binary mixture can favor {\em oblate} uniaxial orientational ordering over {\em prolate}, in sharp contrast with the behavior of the pure systems. This phenomenon gives rise to a new phase diagram topology due to the appearance of {\em two} Landau tetracritical points, leading to a wider region of $N_B$ stability. This feature is shown to hold also for a larger number of components, thus offering an explanation to the results of Ref. \cite{vandenpol}. Finally, we argue that our findings could furnish a new way to look for biaxiality in thermotropic LCs.

\begin{figure}[b]
\center
\includegraphics[scale=0.13]{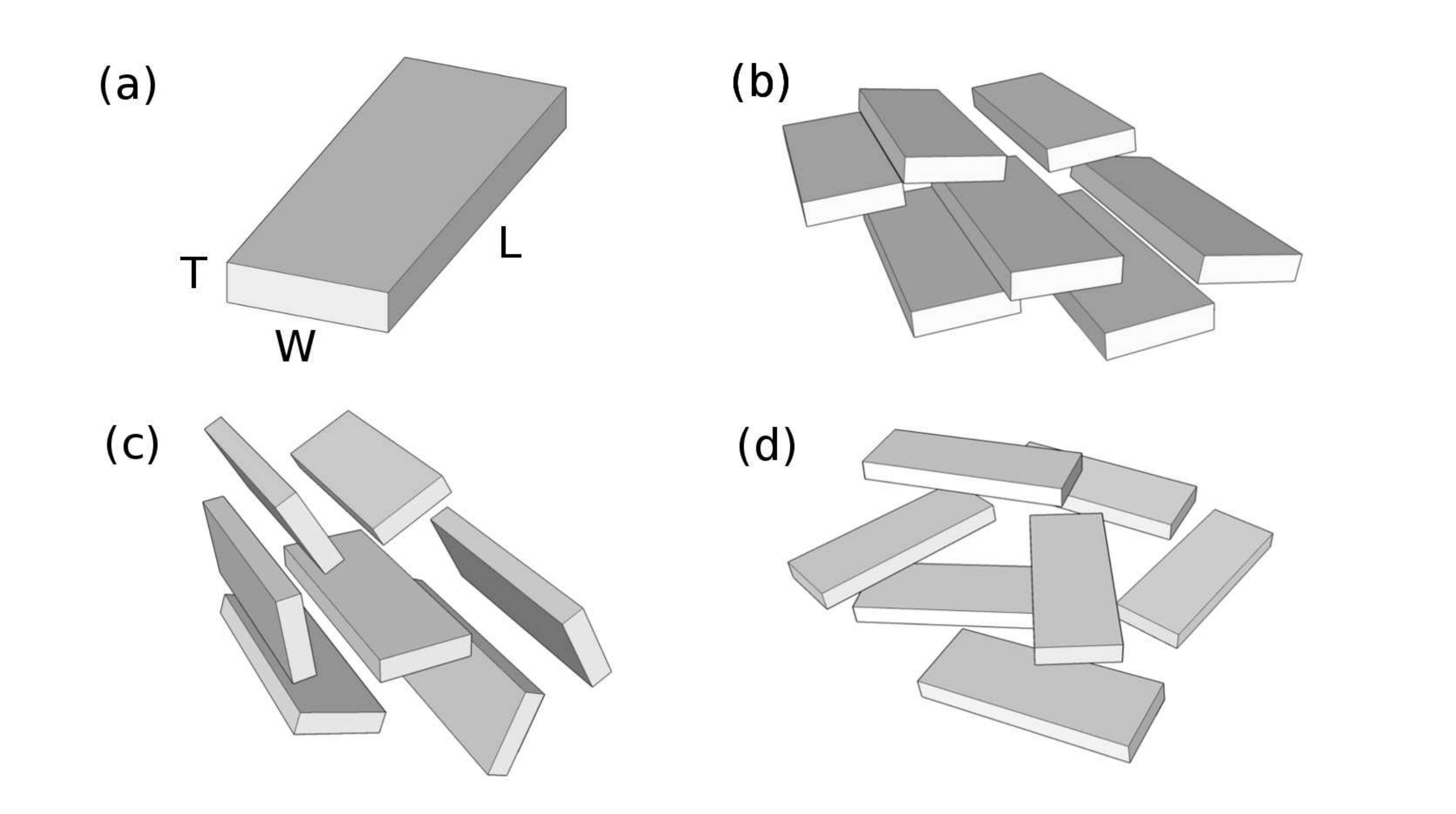}
\caption{\label{f1} (a) Cuboidal particle with dimensions $L\times W \times T$. (b) Schematic representation of a system of freely rotating cuboids in the biaxial nematic phase $N_B$, (c) the uniaxial nematic prolate $N_+$ and (d) the uniaxial nematic oblate $N_-$. In this work the rotational degrees of freedom are discretized according to the Zwanzig model \cite{zwanzig}.} 
\end{figure}

At low density in lyotropics, and at high temperature in thermotropics, the $N_B$ phase appears as a crossover regime in between ``rod-like'' and ``plate-like'' behavior \cite{alben}. In fact, one can distinguish between the $N_U$ phase developed by rods, in which particles align the longest axis along a common direction (uniaxial nematic \textit{prolate}, $N_+$), and that developed by plates, in which particles align the shortest axis (uniaxial nematic \textit{oblate}, $N_-$). A natural candidate system for developing an $N_B$ phase is a binary mixture of rods and plates \cite{vroege}; however, in most cases a demixing transition into two uniaxial nematic phases, i.e. $N_+$ and $N_-$, prevents its stabilization \cite{vanroij, vanderkooij}. Alternatively, a stable $N_B$ state is expected in a system of particles with cuboid (i.e. rectangular parallelepiped) shape defined by the lengths of the principal axes $L \geq W \geq T$, as depicted in Fig. \ref{f1}(a) \cite{straley}. In this case, it is convenient to introduce a \textit{shape parameter} $\nu$, defined by $\nu = \frac{L}{W} - \frac{W}{T}$. By increasing the packing fraction and disregarding the possible stability of inhomogeneous phases, a system of cuboids undergoes an $I \rightarrow N_+ \rightarrow N_B$ sequence of phases if $\nu>0$, whereas an $I \rightarrow N_- \rightarrow N_B$ sequence is found if $\nu<0$ ($I$ stands for the isotropic phase) \cite{mulder}. A schematic representation of these nematic phases is given in Fig. \ref{f1}(b)-(d). The case $\nu=0$ describes the optimal ``brick'' shape exactly in between ``rod-like'' and ``plate-like''. In this case the $N_U$ phase is suppressed and substituted by a second-order $IN_B$ transition \cite{mulder}.

The first experimental realization of the hard-cuboid model was found only recently in a colloidal suspension of boardlike mineral goethite particles \cite{vandenpol}. By producing particles with shape parameter $\nu \simeq 0.1$ close to zero ($\langle L \rangle \times \langle W \rangle \times \langle T \rangle = 254 \times 83 \times 28 \, \mathrm{nm^3}$ and size polydispersity of $20-25\%$), the authors were able to produce an $N_B$ phase stable over a pressure range surprisingly much wider than to be expected from theory \cite{taylor, vanakaras} and simulations \cite{camp} for particles whose shape parameter deviates even slightly from zero. Even more interestingly, the authors affirm that {\em no} $N_U$ phase was observed, contrasting Ref. \cite{mulder}. They suggest that a possible reason for this disagreement should be found in ingredients whose effects have never been studied so far because of their complexity, i.e. fractionation, sedimentation and polydispersity. These unexpected results motivate our interest in analyzing the effect of the above mentioned ingredients, in particular polydispersity, on the stability of the $N_B$ phase in a fluid of hard cuboids.

We consider an $M$-component suspension of $N_\alpha$ colloidal cuboidal particles of species $\alpha=1,...,M$ with dimensions $L_\alpha \times W_\alpha \times T_\alpha$ ($L_\alpha > W_\alpha > T_\alpha$) in a volume $V$ at temperature $T$. The total number density of colloids is $n=\sum_\alpha N_\alpha / V$, the mole fraction of species $\alpha$ is $x_\alpha=N_\alpha/(nV)$ and the packing fraction is $\eta = n \sum_\alpha x_\alpha L_\alpha W_\alpha T_\alpha$. The theoretical framework used in this Letter consists of Onsager theory of LCs \cite{onsager}, which is a density functional theory truncated at second-virial order. In order to facilitate the calculations we follow Zwanzig by restricting the orientations of the particles to the six in which their principal axes are aligned along a fixed Cartesian frame \cite{zwanzig}. Although quantitative agreement with real systems is not expected because of the simplifications introduced in the model, the same model was shown to successfully predict non-trivial phenomena like demixing in rod-plate mixtures \cite{vanroij}, orientational wetting due to confinement and capillary nematization \cite{vanroij2}. Moreover, we expect that transitions between different nematic phases and smectic phases are better described by this model than transitions from isotropic to nematics. In density functional theory the free energy of the system is expressed as a functional of the local density $\rho_i^\alpha(\mathbf{r})$ of particles of species $\alpha=1,...,M$ with orientation $i=1,...,6$ as \cite{evans}

\begin{equation}
\frac{\mathcal{F}[\rho]}{k_B T} = \int d\mathbf{r}\sum_{\alpha,i} \rho_i^\alpha(\mathbf{r})  \Bigl [\ln(\rho_i^\alpha(\mathbf{r}) \Lambda_\alpha^3) -1 \Bigr] + \frac{\mathcal{F}^{ex}[\rho]}{k_B T},
\label{e2}
\end{equation}
where $k_B$ is the Boltzmann constant and $\Lambda_\alpha^3$ the thermal volume of species $\alpha$. At second-virial order the excess free energy $\mathcal{F}^{ex}$ reads

\begin{equation}
\frac{\mathcal{F}^{ex}[\rho]}{k_B T} = - \frac{1}{2} \int d\mathbf{r} \, d\mathbf{r}' \sum_{\alpha,\alpha',i,i'} f_{i i'}^{\alpha \alpha'}({\mathbf r} - {\mathbf r}') \rho_i^\alpha(\mathbf{r}) \rho_{i'}^{\alpha'}(\mathbf{r}'),
\label{e2b} 
\end{equation}
where $f_{i i'}^{\alpha \alpha'}({\mathbf r})=\exp[-u_{i i'}^{\alpha \alpha'}({\mathbf r})/(k_B T)]-1$ is the Mayer function, defined in terms of the pair-wise potential $u_{i i'}^{\alpha \alpha'}({\mathbf r})$. By neglecting spatial modulations, i.e. by imposing $\rho_i^\alpha(\mathbf{r})=\rho_i^\alpha$, the free energy Eq. (\ref{e2}) reduces to an Onsager-type functional whose minimization (under the constraints that $\sum_i\rho_i^\alpha=n x_\alpha$ for all $\alpha=1,...,M$) allows to identify the spatially homogeneous equilibrium phase (see Appendix A). Since at sufficiently high density one expects spatially inhomogeneous phases to be thermodynamically favored, we apply bifurcation theory \cite{mulder} to determine the limit of stability of the homogeneous equilibrium phases with respect to smectic fluctuations. By considering spatial density modulations only along the $z$ axis, i.e $\rho_i^\alpha(\mathbf{r})=\rho_i^\alpha(z)$ in Eq. (\ref{e2}), the smectic bifurcation density is the minimum density at which the Hessian second-derivative matrix of the free energy has an eigenvalue equal to zero (see Appendix B).

Our analysis starts by considering the simplest case of polydispersity, i.e. a mixture of $M=2$ components with mole fractions $x_1$ and $x_2=1-x_1$, respectively. Among the different ways one can parameterize polydispersity, our preliminary analysis suggests to consider \textit{volume polydispersity} (i.e. same particle shape but different volume). Therefore, we study the phase behavior of a binary mixture of hard cuboids whose dimensions are 

\begin{equation}
\begin{array}{c c c}
L_1=L(1+s), & W_1=W(1+s), & T_1=T(1+s), \\
L_2=L(1-s), & W_2=W(1-s), & T_2=T(1-s),
\end{array}
\label{e4}
\end{equation}
where the parameter $s\in[0,1)$ describes the degree of bidispersity. Notice that Eq. (\ref{e4}) implies the same aspect ratios for both species  $L_1/T_1=L_2/T_2=L/T$ and $W_1/T_1=W_2/T_2=W/T$ (hence $\nu_1=\nu_2=\nu$). Here we set $L/T=9.07$ and $W/T=2.96$ ($\nu=0.1$) in order to reproduce the experimental system of Ref. \cite{vandenpol}, thereby neglecting the small effect of the ionic double layer used by the authors to interpret the experimental data.

\begin{figure}
\center
\includegraphics[scale=0.67]{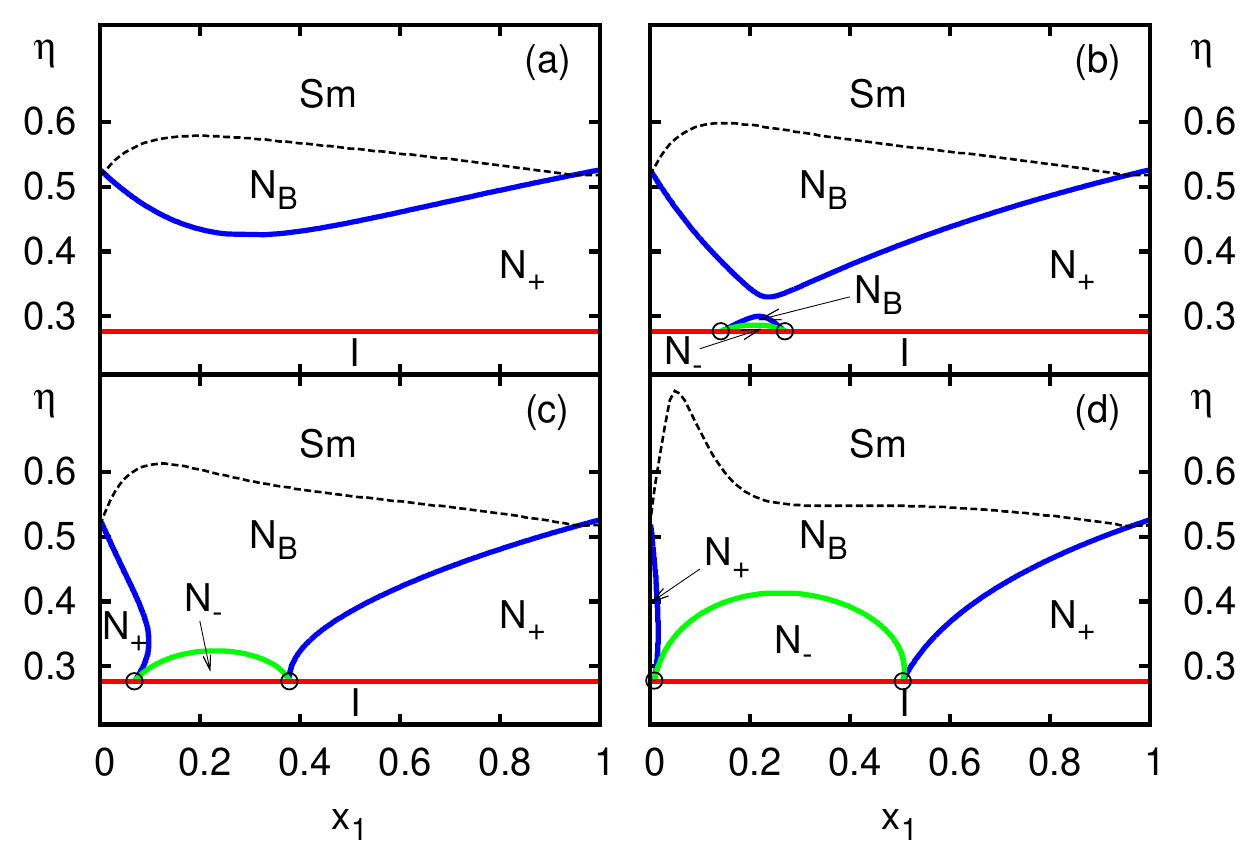}
\caption{\label{f2} (Color online) Phase diagram of a binary mixture of hard cuboids in terms of packing fraction $\eta$ vs. mole fraction of the larger component $x_1$ showing isotropic ($I$), uniaxial ($N_+$ and $N_-$) and biaxial ($N_B$) nematic and smectic ($Sm$) phases. The size of the particles is defined by Eq. (\ref{e4}) with $L/T=9.07$, $W/T=2.96$ ($\nu=0.1$) and bidispersities (a) $s=0.15$, (b) $s=0.18$, (c) $s=0.20$, (d) $s=0.30$. The solid lines separate different homogeneous phases, the dashed lines indicate the limit of stability of the homogeneous phases with respect to smectic fluctuations, whereas the open circles represent the Landau tetracritical points.} 
\end{figure}

Fig. \ref{f2} shows density-composition phase diagrams of binary mixtures ($M=2$) of boardlike particles with the experimental shape parameter $\nu=0.1$ for various bidispersity parameters (a) $s=0.15$, (b) $0.18$, (c) $0.20$ and (d) $0.30$, featuring isotropic ($I$), uniaxial nematic ($N_+$ {\em and} $N_-$), biaxial nematic ($N_B$) and smectic ($Sm$) phases. Due to the near-perfect ``biaxial'' shape of the particles, fractionation is extremely weak and invisible on the scale of Fig. \ref{f2} (see Appendix C). At the extreme mole fractions $x_1=0$ and $x_1=1$ (pure systems) all phase diagrams feature the phase sequence $I\rightarrow N_+ \rightarrow Sm$ that is well known and expected for board-shaped particles with $\nu>0$, with the $N_B$ phase metastable with respect to the $Sm$ phase \cite{mulder,taylor} (see also Appendix D). However, for all $s>0$ there is an intermediate composition regime in which the $N_B$ phase is found to be stable, the more so for increasing $s$. Whereas the opening-up of a stable $N_B$ regime is only quantitative for $s=0.15$, there is a qualitative change of the phase diagram topology beyond $s=0.18$, where {\em two} Landau tetracritical points appear (open circles in Figs. \ref{f2}(b)-(d)). In between these critical points a region of stable $N_-$ phase, which is {\em not} expected for the rod-shaped particles ($\nu>0$) of interest, opens up. Clearly, Figs. \ref{f2}(c) and (d) show that this unexpected $N_-$ regime enlarges with bidispersity, accompanying a further increased $N_B$ stability. In other words, excluded-volume interactions in mixtures of board-shaped rods with the same shape and different volume tend to favor $N_B$ stability as a consequence of an unexpected $N_+ - N_-$ competition. At higher packing fractions the increased $N_B$ stability with respect to the $Sm$ phase is not a surprise, given that regular packing into layers is hindered by size differences between particles \cite{vanakaras}.

\begin{figure}
\center
\includegraphics[scale=0.67]{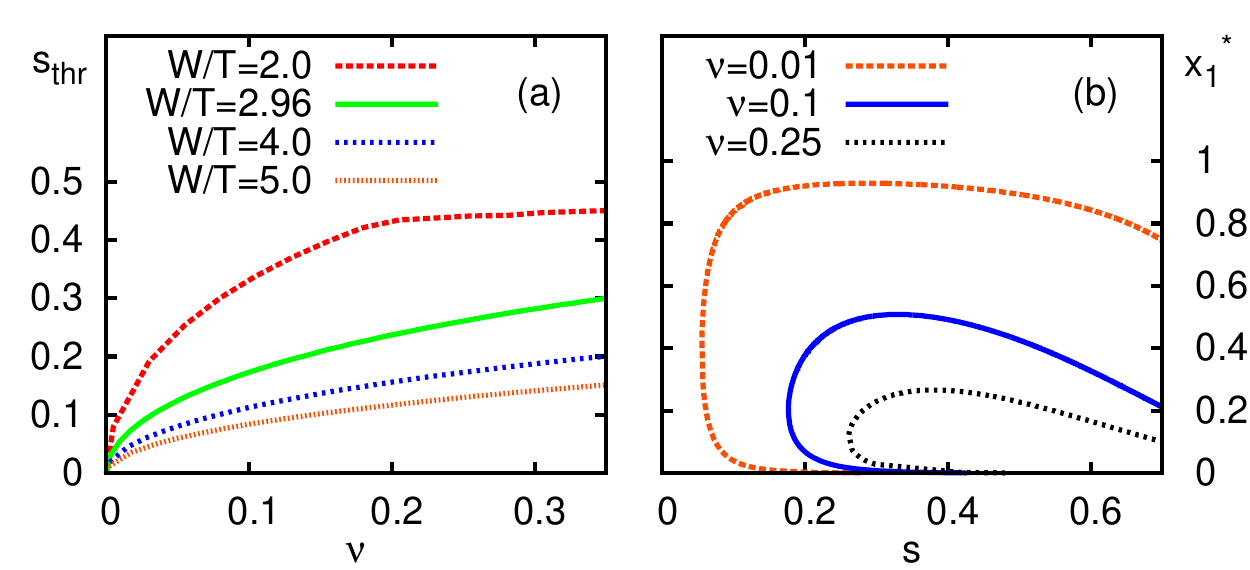}
\caption{\label{f3} (Color online) (a) Threshold bidispersity $s_{thr}$ for the appearance of a tetracritical point as a function of the shape parameter $\nu$ for different fixed values of $W/T$. (b) Critical mole fraction $x_1^*$ as a function of the bidispersity parameter $s$ for a binary mixture of hard cuboids for different shape of the particles (cf. Eq. (\ref{e4})).} 
\end{figure}

It is interesting to understand how the remarkable features of the binary mixture described in Fig. \ref{f2} change with the shape of the particles. Here we are mainly interested in two properties of the phase diagram: (i) the minimum threshold bidispersity $s_{thr}$ at which the Landau tetracritical points appear and (ii) the tetracritical mole fractions $x_1^*$ in terms of the bidispersity $s$. We change the particle shape ($\nu=L/W-W/T$) by fixing in Eq. (\ref{e4}) one aspect ratio ($W/T$) and varying the remaining one ($L/T$). Fig. \ref{f3}(a) shows for $W/T=2.0$, $2.96$, $4.0$ and $5.0$ a similar trend: the minimum threshold bidispersity $s_{thr}$ increases the more the shape deviates from the optimal ``brick'' one. At the same time, the fact that at fixed $\nu$ the threshold bidispersity decreases with $W/T$, indicates that the appearance of the Landau tetracritical points is favored by an increasing aspect ratio of the particles, in qualitative agreement with Ref. \cite{martinez}. Moreover, by fixing the aspect ratio $W/T=2.96$, we can observe the tetracritical mole fraction as a function of the bidispersity for different values of $\nu=0.01$, $0.1$ and $0.25$ in Fig. \ref{f3}(b). The closer the shape is to the optimal ``brick'', the wider is the difference in value of the two tetracritical mole fractions $x_1^*$ and, consequently, the stability regime of the $N_-$ phase. Finally, we note that no critical composition is observed if the particles are closer to the ``plate-like'' shape, i.e. if $\nu_1=\nu_2=\nu<0$ one finds the $N_-$ in between the $I$ and $N_B$ phases for every value of $s$ and $x_1$ (not shown); the $N_+$ phase does {\em not} occur in this case.

\begin{figure}
\center
\includegraphics[scale=0.68]{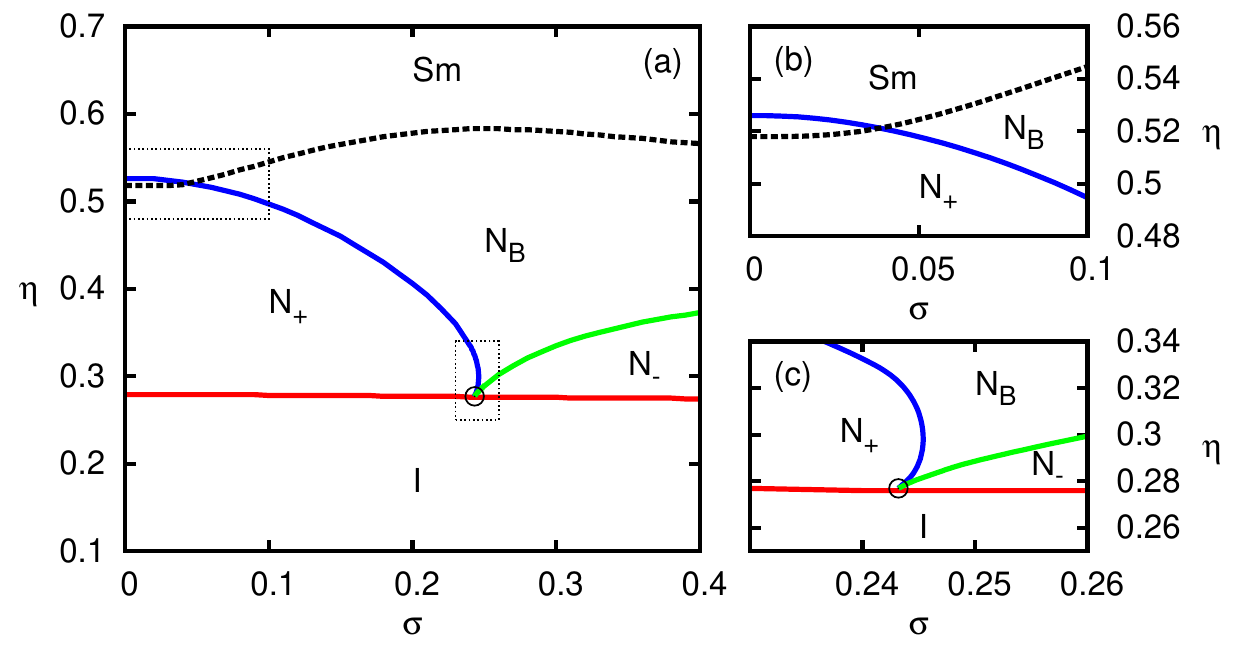}
\caption{\label{f4} (Color online) (a) Phase diagram of $M=21$ components of hard cuboids (packing fraction $\eta$) with aspect ratios $L_\alpha/T_\alpha=9.07$ and $W_\alpha/T_\alpha=2.96$ ($\alpha=1,...,M$) and Gaussian distributed dimensions with polydispersity $\sigma$ (see text). The dashed line indicates the limit of stability of the homogeneous phases with respect to smectic fluctuations. The dotted rectangles highlight (b) the absence of the $N_B$ phase at polydispersity $\sigma<4\%$ due to the direct $N_+ Sm$ phase transition and (c) the reentrant character of the $N_+N_B$ transition close to the tetracritical point (open circle).}
\end{figure}

In order to analyze proper polydispersity, and thus more realistically model the experimental system of Ref. \cite{vandenpol}, we extend our phase-diagram calculations to a system with $M=21$ components of cuboids. Inspired by our analysis of the binary mixture and by the experiments \cite{vandenpol}, we fix the aspect ratios of all species to $L_\alpha/T_\alpha=L/T=9.07$ and $W_\alpha/T_\alpha=W/T=2.96$, such that (i) all species have the same shape $\nu_\alpha=\nu=0.1$ and (ii) the size of each species is completely determined by $T_\alpha$. We consider $T_\alpha$ to be distributed according to a discretized Gaussian function with average $\langle T \rangle =28\,\mathrm{nm}$ and standard deviation $\sigma \langle T \rangle$, where $\sigma$ is the size polydispersity. In general the calculation of a (high-dimensional) phase diagram of a multi-component system is a daunting task \cite{sollich}. In this case, however, it is justified to ignore fractionation (see Appendix C), which reduces the problem to minimizing the functional with respect to $\rho_i^\alpha$ at fixed $n x_\alpha$. The resulting phase diagram in the density-polydispersity representation is shown in Fig. \ref{f4}(a), featuring again $I$, $N_+$, $N_-$, $N_B$ and $Sm$ equilibrium states and a tetracritical point at $\sigma \simeq 24 \%$, which is surprisingly close to the size polydispersity in the experiments \cite{vandenpol}. The strikingly large stability regime of the $N_B$ is caused by the reduced stability of $Sm$ and $N_+$ (cf. Fig. \ref{f4}(b)), not unlike in the binary case. However, a direct $I N_B$ transition similar to that observed in Ref. \cite{vandenpol} is not expected in this model due to the reentrant character of the $N_+ N_B$ phase transition (cf. Fig. \ref{f4}(c)).     

In conclusion, by means of a mean-field theoretical approach with discrete orientations we have shown that size polydispersity strongly affects the phase behavior of boardlike particles, driving the emergence of a novel topology of the phase diagram. This topology change is due to the appearance of Landau tetracritical points, which in turn is related to a competition between the prolate ``rod-like'' ordering typical of the pure components and the oblate ``plate-like'' purely induced by the mixing. In combination with the destabilization of the $Sm$ phase, we can conclude that polydispersity dramatically increases the stability regime of the $N_B$ phase. The usual stability limitations of $N_B$ phases, such as $N_+ - N_-$ demixing of rod-plate mixtures and ordering into smectics, are therefore overcome in the present system. Although this work focuses on a particular value of the particles dimensions, its predictions hold for a more general choice of the relevant parameters, as reported in Fig. \ref{f3}. Moreover, we do not expect the homogeneous phase behavior to be crucially dependent on the form of the interaction (cuboidal), on the contrary it should be qualitatively similar to other excluded-volume interactions with the {\em same} symmetry (e.g. spheroid, spheroplatelet).
 
Finally, it is tempting to consider this work in the perspective of stabilizing $N_B$ thermotropic liquid crystals. In this case, the soft-core character of the inter-molecular interactions does not allow for a univocal definition of ``shape'', and van der Waals forces can significantly influence the phase diagram. Nonetheless, it is widely accepted that hard-core models contain the essential physical ingredients for a first-approximation description of the structure of a molecular or colloidal fluid \cite{frenkel}. Following this interpretation scheme, it is intriguing to wonder whether it is possible to enhance the $N_B$ stability by considering two- or multi-component mixtures of molecules with biaxial symmetry and different size. We hope our findings will stimulate further research in this direction.

\smallskip   

This work is financed by a NWO-VICI grant and is part of the research program of FOM, which is financially supported by NWO.

\appendix

\section{Density functional theory}\label{dft}

In the present work the orientational degrees of freedom of the particles are treated within the Zwanzig model \cite{zwanzig}, hence a particular orientation can be identified with a number $i=1,...,6$ (cf. Tab. \ref{tab1}).

\begin{table}[htp]
 \centering

\begin{center}
\begin{tabular} {|c||c|c|c|}
\hline
\hspace*{0.5cm}\textit{i}\hspace*{0.5cm} & \hspace*{0.5cm}\textit{L}\hspace*{0.5cm} & \hspace*{0.5cm}\textit{W}\hspace*{0.5cm} & \hspace*{0.5cm}\textit{T}\hspace*{0.5cm}\\ \hline \hline
1 & x & y & z\\ \hline
2 & z & x & y\\ \hline
3 & y & z & x\\  \hline
4 & x & z & y\\ \hline
5 & y & x & z\\ \hline
6 & z & y & x\\ 
\hline
\end{tabular}
\end{center}
 
 \caption{Enumeration of the orientational configurations of a hard cuboid within the Zwanzig model. Each configuration $i$ is identified with the directions ($x,y,z$) along which the particle axes ($L,W,T$) are aligned.}
 \label{tab1}
\end{table}

According to density functional theory it is possible to express the free energy of a system as a functional of the single-particle density $\rho_i^\alpha(\mathbf{r})$ of particles with orientation $i$ ($i=1,...,6$) belonging to species $\alpha$ ($\alpha=1,...,M$) as \cite{evans} 

\begin{equation}
\frac{\mathcal{F}[\rho]}{k_B T} = \int d\mathbf{r}\sum_{\alpha,i} \rho_i^\alpha(\mathbf{r})  \Bigl [\ln(\rho_i^\alpha(\mathbf{r}) \Lambda_\alpha^3) -1 \Bigr] + \frac{\mathcal{F}^{ex}[\rho]}{k_B T},
\label{eq2}
\end{equation}
where for brevity

\begin{equation*}
 \sum_i \equiv \sum_{i=1}^6, \qquad \sum_\alpha \equiv \sum_{\alpha=1}^M, \qquad  \int d \mathbf r \equiv \int_V d \mathbf r. 
\end{equation*}
The excess term $\mathcal{F}^{ex}[\rho]$ has in general a non-trivial dependence on $\rho_i^\alpha(\mathbf{r})$. For short-range potentials it is always possible to express $\mathcal{F}^{ex}[\rho]$ as a virial series in the single-particle density. Therefore, by truncating the series at second-virial order and thus disregarding higher-order contributions, one obtains

\begin{equation}
\frac{\mathcal{F}^{ex}[\rho]}{k_B T} = - \frac{1}{2} \int d\mathbf{r} \, d\mathbf{r}' \sum_{\alpha,\alpha',i,i'} f_{i i'}^{\alpha \alpha'}({\mathbf r} - {\mathbf r}') \rho_i^\alpha(\mathbf{r}) \rho_{i'}^{\alpha'}(\mathbf{r}'),
\label{eq2b}
\end{equation}
where the Mayer function $f_{i i'}^{\alpha \alpha'}({\mathbf r})$ is defined in terms of the pairwise interaction potential $u_{i i'}^{\alpha \alpha'}({\mathbf r})$ as

\begin{equation}
f_{i i'}^{\alpha \alpha'}({\mathbf r}) = \exp \biggl [-\frac{u_{i i'}^{\alpha \alpha'}({\mathbf r})}{k_B T} \biggr ] -1.
\label{eq3}
\end{equation}
The single-particle density $ \rho_i^\alpha(\mathbf r)$ is related to the number of particles $N_\alpha$ through the normalization condition

\begin{equation}
\int d \mathbf r \sum_i \rho_i^\alpha(\mathbf r) = N_\alpha = x_\alpha N. 
\label{eq4}
\end{equation}

For hard cuboids the interaction potential, which expresses the impenetrability of the particles, is 

\begin{equation}
\frac{u_{i i'}^{\alpha \alpha'}(\mathbf{r})}{k_B T} =
\begin{cases}
\infty \hspace{0.8cm} \text{if $|x|<(X^{\alpha}_{i}+X^{\alpha'}_{i'})$ } \\
\hspace{1.15 cm} \text {and $|y|<(Y^{\alpha}_{i}+Y^{\alpha'}_{i'})$} \\
\vspace{0.15 cm}
\hspace{1.15 cm} \text{and $|z|<(Z^{\alpha}_{i}+Z^{\alpha'}_{i'})$;}\\
0 \hspace{1cm} \text{otherwise.}
\end{cases} 
\label{eq5}
\end{equation}
According to the index notation defined in Tab. \ref{tab1}, the $6$-dimensional vectors $\mathbf{X}^{\alpha}$, $\mathbf{Y}^{\alpha}$ and $\mathbf{Z}^{\alpha}$ of species $\alpha$ are given in terms of the dimension of the particles by

\begin{equation}
\begin{array}{c}
\vspace{0.15 cm}
\mathbf{X}^{\alpha} = \frac{1}{2}(L_\alpha,W_\alpha,T_\alpha,L_\alpha,W_\alpha,T_\alpha), \\
\vspace{0.15 cm}
\mathbf{Y}^{\alpha} = \frac{1}{2}(W_\alpha,T_\alpha,L_\alpha,T_\alpha,L_\alpha,W_\alpha), \\
\mathbf{Z}^{\alpha} = \frac{1}{2}(T_\alpha,L_\alpha,W_\alpha,W_\alpha,T_\alpha,L_\alpha).
\end{array}
\label{eq6}
\end{equation}

The main goal of this work is to study the stability of spatially homogeneous phases (i.e. isotropic and nematic). In order to simplify the problem we therefore neglect spatial modulations in the single-particle density by imposing $\rho_i^\alpha(\mathbf{r})=\rho_i^\alpha$. Consequently, Eq. (\ref{eq2}) becomes

\begin{equation}
\frac{\mathcal{F}}{V k_B T} = \sum_{\alpha,i} \rho_i^\alpha  \Bigl [\ln(\rho_i^\alpha \Lambda_\alpha^3) -1 \Bigr] + \frac{1}{2} \sum_{\alpha,\alpha',i,i'} E^{\alpha \alpha'}_{i i'} \rho_i^\alpha \rho_{i'}^{\alpha'},
\label{eq7}
\end{equation}
which is the restricted orientation version of the Onsager free energy \cite{onsager}. The matrix $E^{\alpha \alpha'}_{i i'}$ in Eq. (\ref{eq7}) is the excluded volume between two particles belonging to species $\alpha$ and $\alpha'$ with orientations $i$ and $i'$ interacting through the potential Eq. (\ref{eq5})

\begin{equation}
E_{i i'}^{\alpha \alpha '} = 8(X^{\alpha}_{i}+X^{\alpha'}_{i'})(Y^{\alpha}_{i}+Y^{\alpha'}_{i'})(Z^{\alpha}_{i}+Z^{\alpha'}_{i'}). 
\label{eq9}
\end{equation}
In the homogeneous case the normalization condition Eq. (\ref{eq4}) becomes

\begin{equation}
 \sum_i \rho_i^\alpha = x_\alpha n.
\label{eq8}
\end{equation}
The single-particle density at equilibrium is the one which minimizes Eq. (\ref{eq7}) with the constraints of Eq. (\ref{eq9}) for all $\alpha=1,...,M$, hence it is found by solving the Euler-Lagrange equation

\begin{equation}
\rho_i^\alpha = \frac{x_\alpha n \exp \biggl ( - \frac{1}{2} \sum_{\alpha', i'} E_{i i'}^{\alpha \alpha '} \rho_{i'}^{\alpha'} \biggr)}{\sum_{i''}\exp \biggl ( - \frac{1}{2} \sum_{\alpha', i'} E_{i'' i'}^{\alpha \alpha '} \rho_{i'}^{\alpha'} \biggr)},
\label{eq9b}
\end{equation}
which is achieved by standard numerical (iterative) techniques. 

\section{Nematic-Smectic bifurcation}\label{bif}

While studying the homogeneous equilibrium phases of the system, we are also interested in estimating their upper bound in the phase diagram, where spatially inhomogeneous phases tend to be thermodynamically favored. Bifurcation theory \cite{kaiser, mulder2} provides a way to investigate the limit of stability of a particular phase. 

The condition of thermodynamic stability of a phase described by the single-particle density $\rho_i^\alpha(\mathbf r)$ requires that the system corresponds to a minimum of the free energy $\mathcal{F}$, i.e. a stationary point that satisfies

\begin{equation}
\int d\mathbf{r} \, d\mathbf{r}' \sum_{\alpha, \alpha',i, i'} \frac{\delta^2 F}{\delta \rho_i^\alpha(\mathbf{r}) \delta \rho_{i'}^{\alpha'}(\mathbf{r}')} \delta \rho_i^\alpha(\mathbf{r}) \delta \rho_{i'}^{\alpha'}(\mathbf{r}') >0,
\label{eq10} 
\end{equation} 
for any arbitrary perturbation $\delta \rho_i^\alpha(\mathbf r)$. By inserting the functional expression Eq. (\ref{eq2}) into Eq. (\ref{eq10}), one finds that the reference phase (described by $\rho_i^\alpha(\mathbf r)$) ceases to be stable at the smallest density $n=N/V$ at which a perturbation $\delta \rho_i^\alpha(\mathbf r)$ exists such that
 
\begin{equation}
\delta \rho_i^\alpha(\mathbf r) = \rho_i^\alpha(\mathbf r) \int d \mathbf r' \sum_{\alpha',i'} f_{i i'}^{\alpha \alpha'}(\mathbf r - \mathbf r') \delta \rho_{i'}^{\alpha'}(\mathbf r').
\label{eq11}
\end{equation}

Here we are interested in calculating the limit of stability of the (uniaxial or biaxial) nematic phase with respect to smectic fluctuations. With this in mind, in Eq. (\ref{eq11}) we neglect spatial modulations in the reference phase, i.e. $\rho^\alpha_i(\mathbf{r})=\rho^\alpha_i$, and a positional dependence of the fluctuations only along the $z$ direction, i.e. $\delta \rho^\alpha_i(\mathbf{r})=\delta \rho^\alpha_i(z)$. After some rearranging Eq. (\ref{eq11}) becomes

\begin{equation}
\sigma^\alpha_i(z) = \sum_{\alpha', i'} \int dz' \, Q_{ii'}^{\alpha \alpha'}(z-z') \, \sigma^{\alpha'}_{i'}(z'),
\label{eq24}
\end{equation}
where $\sigma^\alpha_i(z)=\delta\rho^\alpha_i(z)/\sqrt{\rho^\alpha_i}$ and

\begin{equation}
Q_{ii'}^{\alpha \alpha'}(z)= \sqrt{\rho^\alpha_i \rho^{\alpha'}_{i'}} \int dx \, dy \, f_{ii'}^{\alpha \alpha'}(\mathbf{r}),
\label{eq25}
\end{equation}
a symmetric (Hermitean) kernel.
By inserting the explicit form of the inter-particle potential (cf. Eq. (\ref{eq3}) and (\ref{eq5})) into Eq. (\ref{eq25}), we obtain

\begin{equation}
Q_{ii'}^{\alpha \alpha'}(z) =
\begin{cases}
-4\sqrt{\rho^\alpha_i \rho^{\alpha'}_{i'}}(X^{\alpha}_{i}+X^{\alpha'}_{i'})(Y^{\alpha}_{i}+Y^{\alpha'}_{i'}) \hspace{0.8cm} \\
\vspace{0.2 cm}
\hspace{1.85cm} \text{if $|z|<(Z^{\alpha}_{i}+Z^{\alpha'}_{i'})$;} \\
0 \hspace{1.7cm} \text{otherwise.}
\end{cases} 
\label{eq26}
\end{equation}
Eq. (\ref{eq24}) can be more conveniently solved in Fourier space, where it reads

\begin{equation}
\hat{\sigma}^\alpha_i(q) = \sum_{\alpha', i'} \hat{Q}_{ii'}^{\alpha \alpha'}(q) \hat{\sigma}^{\alpha'}_{i'}(q),
\label{eq27}
\end{equation}
with

\begin{equation}
\hat{Q}_{ii'}^{\alpha \alpha'}(q) = - \sqrt{\rho^\alpha_i \rho^{\alpha'}_{i'}} E_{ii'}^{\alpha \alpha'}j_0\bigl (q(Z^{\alpha}_{i}+Z^{\alpha'}_{i'})\bigr),
\label{eq28}
\end{equation}
and $j_0(x)=\sin(x)/x$.

In conclusion, the limit of stability of the nematic phase with respect to smectic fluctuations can be numerically found as the minimum packing fraction $\eta^*$ at which there exists a wave vector $q^*$ such that the $6M \times 6M$ matrix with entries $\hat{Q}_{ii'}^{\alpha \alpha'}(q^*)$ has a unit eigenvalue. The periodicity of the corresponding bifurcating smectic phase is given by $\lambda^*=2\pi/q^*$.

\section{Nearly second-order character of the $I N_{\pm}$ transition}

When dealing with mixtures, the phase diagram is conveniently expressed in terms of the osmotic pressure $P$ vs. the mole fraction $x_\alpha$ of $M-1$ components. In this way it is possible to visualize the coexistence of phases characterized by a different composition with respect to the parent distribution. This phenomenon, called demixing or fractionation, is a consequence of the first-order character of the transition. 

\begin{figure}
\center
\includegraphics[scale=0.67]{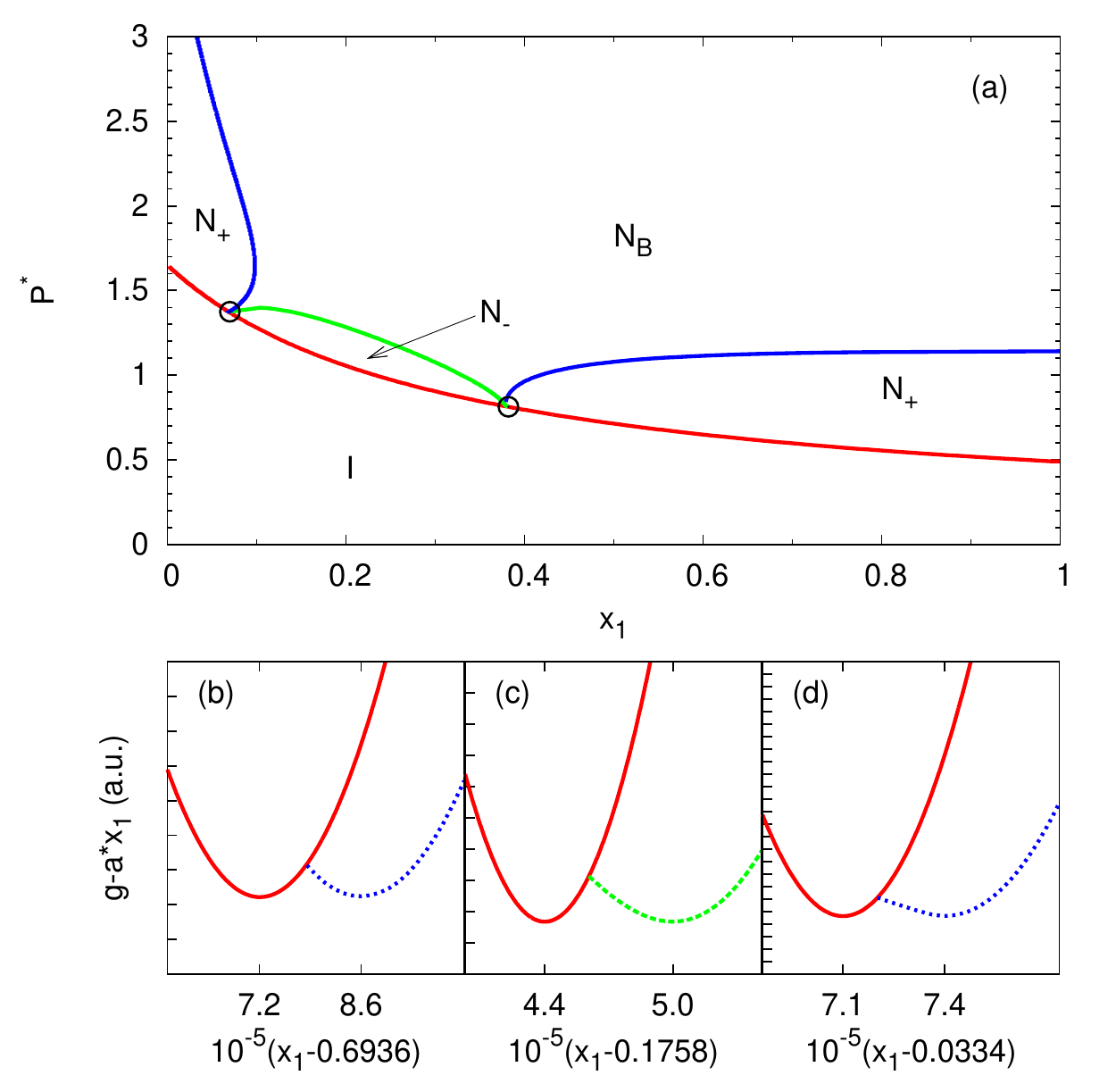}
\caption{\label{fig3} (a) Phase diagram of a binary mixture of hard cuboids in terms of the reduced osmotic pressure $P^*=PLWT/(k_BT)$ vs. mole fraction of the first species $x_1$. (b)-(d) $I$ (red solid line), $N_+$ (blue dotted line) and $N_-$ (green dashed line) branches of the Gibbs free energy per particle $g=G/N$ at (b) $P^*=0.6$, (c) $P^*=1.1$ and (d) $P^*=1.5$. A straight line with slope $a=\partial g / \partial x_1|_{x_1=x_I}=\partial g / \partial x_1|_{x_1=x_N}$ (with $x_I$ and $x_N$ the compositions of the coexisting isotropic and nematic phases) was subtracted in each case to enhance the visualization of the common tangent construction.} 
\end{figure}

Here we analyze demixing in a binary mixture of cuboids parameterized as in Eq. (3) with $L/T=9.07$, $W/T=2.96$ and $s=0.2$. In Fig. \ref{fig3}(a) we report the phase diagram for such a system as a function of the mole fraction $x_1$ of the larger species. The expected first-order character of the $I N_U$ transitions is not detectable at this scale (see below), whereas the $N_U N_B$ transitions appear to be second order. At three different values of the reduced pressure $P^*=P L W T / (k_B T)$ we calculated the isotropic and uniaxial nematic branches of the Gibbs free energy per particle $g(P,x_1)=G(P,N_1,N_2)/(k_B T(N_1+N_2))$. The coexistence between the two phases is given by a common tangent construction, which allows to evaluate the difference in composition $\Delta x_1$ of the coexisting phases. The results are reported in Fig. \ref{fig3}(b)-(d) for $P^*=0.6$, $1.1$ and $1.5$, respectively. In the three cases, two of which describe a $I N_+$ and one a $I N_-$ transition, $\Delta x_1 \approx 10^{-5}$ and can therefore be neglected. The situation does not change when one considers different values of the bidispersity parameter $s$.

Although Landau-de Gennes theory predicts the $I N_U$ transition to be first order \cite{degennes}, we have just shown that its discontinuous character can be safely neglected for the binary mixture of boardlike particles we consider in this work. In our opinion, this fact is tightly related to the shape of the particles close to the $\nu=0$ value. In fact, when considering  a monodisperse system, the closer $\nu$ is to zero the weaker is the first-order character of the $I N_U$ transition (see also Sec. \ref{mono}). This fact allows us to assume that for an arbitrary number of components of volume-polydisperse cuboids close to $\nu=0$ the $I N_U$ transition can be approximated as continuous. As a consequence, we can neglect demixing in the phase behavior analysis reported in Fig. 4, thus reducing enormously the complexity of the problem.

\section{Monodisperse system of hard cuboids}\label{mono}

The main goal of the present work is to investigate how polydispersity affects the phase behavior, and in particular the stability, of the $N_B$ phase in a system of hard cuboids. For this reason, it is instructive to study what the theoretical framework described in Sec. \ref{dft} predicts in the monodisperse case $M=1$. In particular, we will focus here on the role of the particles dimensions on the phase behavior of the system.

\begin{figure}
\center
\includegraphics[scale=0.67]{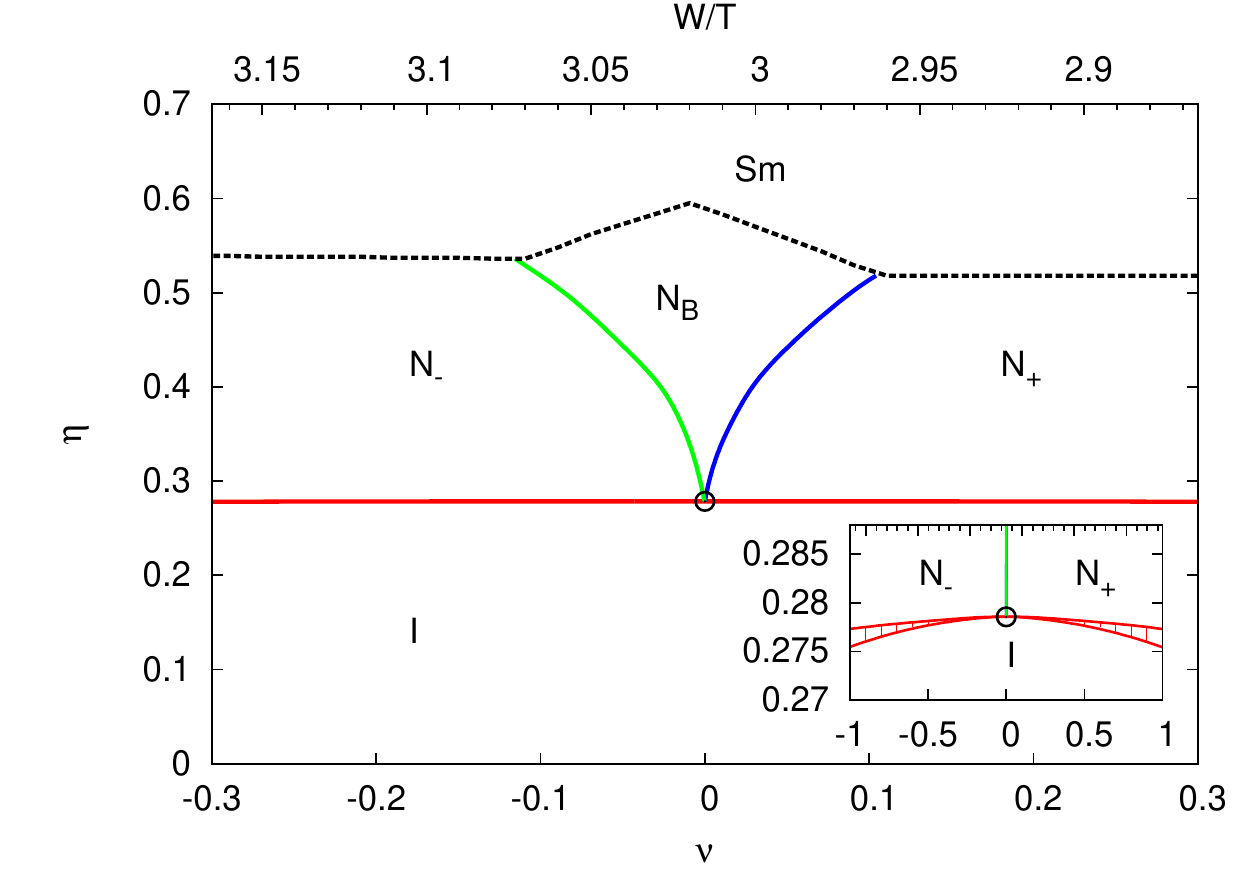}
\caption{\label{fig1} Phase diagram of a monodisperse system of hard cuboids as a function of the shape parameter $\nu=L/W-W/T$ (with $L/T=9.07$ fixed and $W/T$ variable). The solid lines indicate phase boundaries as calculated by minimizing the Onsager-Zwanzig functional, the dashed line indicates the limit of stability of the nematic with respect to the smectic phase and the open circle the Landau tetracritical point. The inset highlights the first order character of the $I N_U$ transition and how this tends to become continuous by approaching $\nu=0$.} 
\end{figure}

In Fig. \ref{fig1} we report the phase diagram of a monodisperse system of hard cuboids as a function of the aspect ratio $W/T$ at fixed $L/T=9.07$. Consequently, by varying $W/T$ one varies the shape parameter $\nu=L/W-W/T$, in such a way that by crossing the point $\nu=0$ one expects a transition from plate- to rod-like behavior. This is precisely what Fig. \ref{fig1} shows, where the phase separation lines are calculated by minimizing the Onsager-Zwanzig functional Eq. (\ref{eq7}) with the constraint of Eq. (\ref{eq8}) for each value of the packing fraction $\eta$. Moreover, bifurcation theory (cf. Sec. \ref{bif}) provides a way to estimate the upper limit of stability of homogeneous phases with respect to the smectic (dashed line in Fig. \ref{fig1}). Fig. \ref{fig1} shows that to observe a stable $N_B$ phase, the shape of the particles should be designed with extremely high precision in a small $\nu$-regime about $\nu=0$. In fact, for $L=9.07 \, T$ the $N_B$ phase disappears unless $2.96 \, T<W<3.08 \, T$. This is due both to the tight cusp-like shape of the $N_U N_B$ transition line and to the preempting character of inhomogeneous phases. Analogous results can be obtained by varying the shape parameter through $L/T$, while keeping $W/T$ fixed (not shown). Finally, in the inset of Fig. \ref{fig1} (note the different scale) we show the first order character of the $I N_U$ transition, which tends to become second-order by approaching the critical point at $\nu=0$.

\begin{figure}
\center
\includegraphics[scale=0.67]{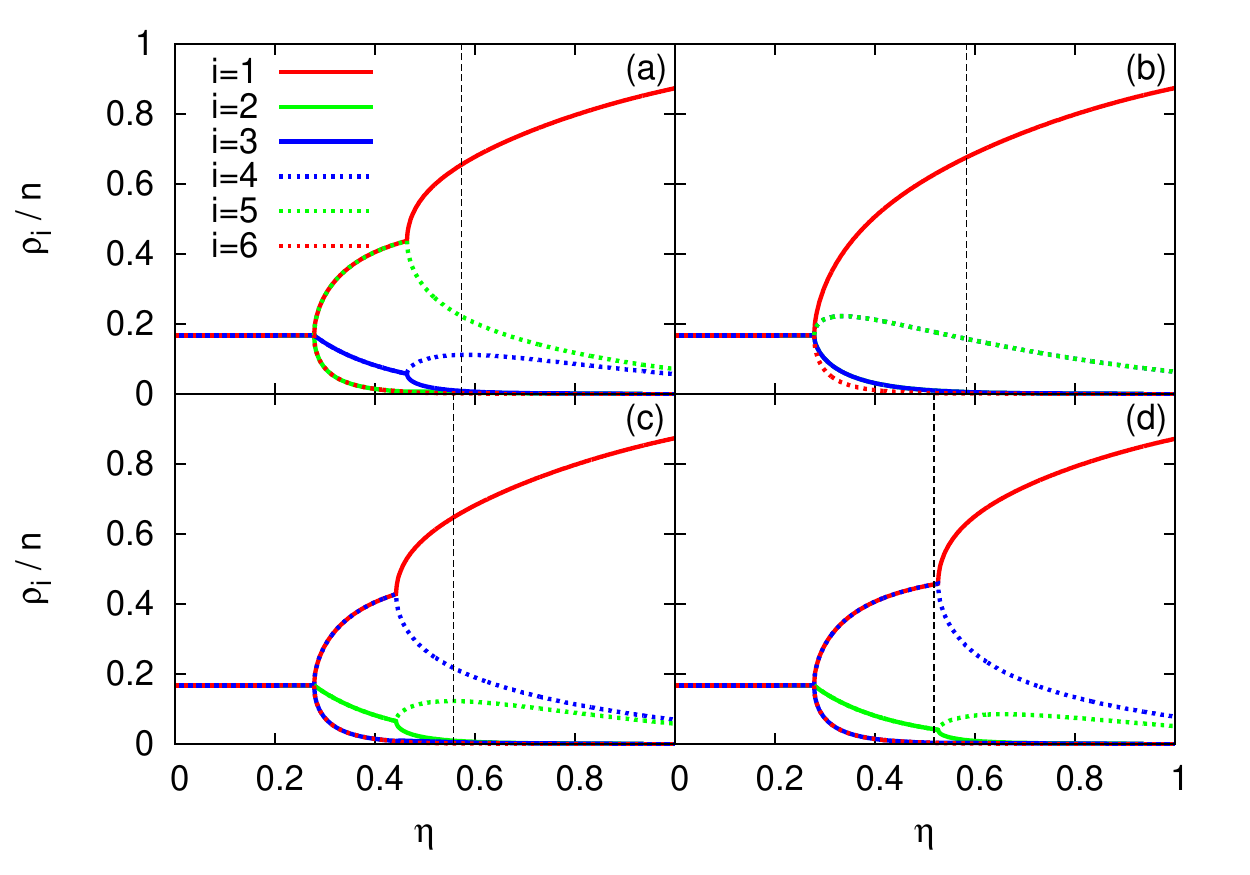}
\caption{\label{fig2} Orientation distribution function of a monodisperse system of hard cuboids as a function of the packing fraction $\eta$ obtained by minimization of the Onsager-Zwanzig functional Eq. (\ref{eq7}) for $M=1$. The cuboids have dimensions $L/T=9.07$ and (a) $W/T=3.04$, (b) $W/T=3.01$, (c) $W/T=2.99$, (d) $W/T=2.96$. The different lines indicate the probability of a particular orientation $i=1,...,6$ (cf. Tab. \ref{tab1}). The dashed vertical line shows the limit of stability of the nematic phases with respect to the smectic.} 
\end{figure}

For the sake of completeness, in Fig. \ref{fig2} we report the orientation distribution function $p_i$, which is the probability of a given orientation $i=1,...,6$ as a function of the packing fraction $\eta$ for different values of the shape parameter $\nu$. In the monodisperse case this function coincides with the single-particle density divided by the number density: $p_i=\rho_i/n$. The values of the orientation distribution function characterize the symmetry of the corresponding phase. In fact, at a given packing fraction $\eta$ in Fig. \ref{fig2} one can have one of the following possibilities:

\begin{itemize}
\item the probabilities $p_i$ are all the same, i.e. $p_i=1/6$ (isotropic $I$ phase);
\item the probabilities $p_i$ are coupled two-by-two, demonstrating the presence of a symmetry axis (uniaxial nematic $N_U$ phase);
\item the probabilities $p_i$ are different between each others (biaxial nematic $N_B$ phase).
\end{itemize}
Moreover, in the uniaxial nematic case one can further distinguish two situations:

\begin{list}{$\blacktriangle$}{}
\item the two more probable orientations have the shortest axis aligned along the same direction (uniaxial nematic oblate $N_-$ phase);
\item the two more probable orientations have the longest axis aligned along the same direction (uniaxial nematic prolate $N_+$ phase).
\end{list}
This classification is easily generalized to the multi-component case. With this in mind, one can observe the difference in the orientation distribution function when $\nu=-0.06<0$ ($W/T=3.04$, Fig. \ref{fig2}(a)), $\nu=0$ ($W/T=3.01$, Fig. \ref{fig2}(b)) and $\nu=0.04>0$ ($W/T=2.99$, Fig. \ref{fig2}(c)). The vertical dashed line indicates the limit of stability with respect to smectic fluctuations as given by bifurcation theory. Finally, Fig. \ref{fig2}(d) shows the predicted orientation distribution function when the experimental value $W/T=2.96$ is considered \cite{vandenpol}, and highlights how according to the model the $N_B$ phase is expected to be preempted by inhomogeneous phases.

\end{document}